\documentstyle[12pt]{article}
\def\selecmp{\tilde e^{\mp}}

\def\selecR{\tilde e_{\rm R}}
\def\selecL{\tilde e_{\rm L}}

\def\lsp{{\tilde\chi^0}_1}
\def\photino{\tilde\gamma}
\def\charglm{\tilde {w_{\rm 1}}^-}
\def\charglp{\tilde {w_{\rm 1}}^+}

\begin{document}
\pagestyle{plain}
\renewcommand{\thefootnote}{*)}

\title{\bf
\begin{quote}
\raggedleft TMCP--95--2\\
March 1995\\ ~
\end{quote}
The automatic computation for SUSY processes}
\font\fonts=cmbx12
\author{~\\
\bf Masato J{\fonts IMBO}\\
\it Computer Science Laboratory, Tokyo Management College,\\
\it Ichikawa, Chiba 272, Japan\\
\it (e-mail: jimbo@kekvax.kek.jp)\\
\and \bf Tadashi K{\fonts ON}\\
\it Faculty of Engineering, Seikei University,\\
\it Musashino, Tokyo 180, Japan\\
\it (e-mail: kon@ge.seikei.ac.jp)\\
\and \bf M{\fonts INAMI}-T{\fonts ATEYA} collaboration\\
}\date{ }
\maketitle
\begin{abstract}
    We have constructed a system for the automatic
computation of cross-sections for the processes of the
SUSY QED by the extension of the GRACE system including
a Majorana fermion.   The system has also been applied to
another model including Majorana fermions, MSSM, by the
definition of the model file.
\end{abstract}
\renewcommand{\thefootnote}{\sharp\arabic{footnote}}
\renewcommand{\theequation}{\arabic{section}.\arabic{equation}}
\setcounter{footnote}{0}

\quad\\
\noindent{\bf Introduction}

    It has been widely believed that there exists a symmetry called
supersymmetry (SUSY) between bosons and fermions at the unification-energy
scale.  It, however, is a broken symmetry at the electroweak-energy scale.
The relic of SUSY is expected to remain as a rich spectrum of SUSY particles,
partners of usual matter fermions, gauge bosons and Higgs scalars, named
sfermions, gauginos and higgsinos, respectively~\cite{theor-a,theor-b,theor-c}.

    The neutral gauginos and higgsinos are Majorana fermions, which become
the mixed states called neutralinos.  Since anti-particles of Majorana fermions
are themselves, there exists so-called `Majorana-flip', the transition between
particle and anti-particle.  This has been the most important problem which we
should solve when we realize the automatic system for computation of the SUSY
processes.

    In a recent work~\cite{jt,jtkk}, we developed an algorithm to treat
Majorana fermions in the program package CHANEL~\cite{chan} which has been
developed for the numerical calculation of the helicity amplitudes.  We have
already possessed the GRACE system~\cite{pre-grc} which has been developed for
the computation of the matrix elements for the processes of the standard model.
The GRACE system automatically generates the source code for CHANEL, and
includes the interface and library of CHANEL, and the multi-dimensional
integration package BASES~\cite{bas}.

    In the standard model, we already have such particles as Dirac fermions,
gauge bosons and scalar bosons in the GRACE system.  Thus we can construct
an automatic system for the computation of the SUSY processes by the algorithm
above in the GRACE system.  In this work, we present the check list of the
system.

\quad\\
\noindent{\bf Majorana fermions into new GRACE}

   The method of computation in the program package CHANEL is as follows:
\begin{enumerate}
  \item To divide a helicity amplitude into vertex amplitudes.
  \item To calculate each vertex amplitude numerically as a complex number.
  \item To reconstruct of them with the polarization sum, and calculate
  the helicity amplitudes numerically.
\end{enumerate}
The merit of this method is that the extension of the package is easy,
and that each vertex can be defined only by the type of concerned particles.

    When we adopt the algorithm in Ref.~\cite{jt,jtkk} for the implementation
of the embedding Majorana fermions in CHANEL, the kinds of the
Dirac-Majorana-scalar vertices are limited to four types:
\begin{itemize}
\begin{itemize}
  \item[(1)] $\overline{U} \Gamma U$
  \item[(2)] $U^{\rm T} \Gamma \overline{U}~^{\rm T}$
  \item[(3)] $\overline{U} C^{\rm T} \Gamma^{\rm T} \overline{U}~^{\rm T}$
  \item[(4)] $U^{\rm T} \Gamma^{\rm T} C^{-1} U$
\end{itemize}
\end{itemize}
where $U$'s denote wave functions symbolically without their indices, and $C$
is the charge-conjugation matrix. The symbol $\Gamma$ stands for the scalar
vertex such as
\[ \Gamma = A_{\rm L}\cdot{{1 - \gamma}\over{2}} +
A_{\rm R}\cdot{{1 + \gamma}\over{2}} ~~. \]
The vertices (2)$\sim$(3) are related to the vertex (1) which is the same
as the Dirac-Dirac-scalar vertex in the subroutine of CHANEL.  Thus we can
build three new subroutines for the added vertices.

    On the other hand, the GRACE system becomes more flexible for the extension
in the new version called `{\bf grc}'~\cite{grc-pp}, which includes a new
graph-generation package.  With this package, graphs can be generated based on
a user-defined model.  We have performed the installation of the subroutines
above with the interface on the new GRACE system.

\quad\\
\noindent{\bf Results for tests}

    At the start for the check of our system, we have written the model file
of SUSY QED.  In this case, there is only one Majorana fermion, photino.
Next we have extended the model file and the definition file of couplings for
MSSM.  The tests have been performed by the exact calculations with the two
methods, our system and REDUCE.  In Table I, the tested processes are shown as
a list.  The references in the table are not the results of the tests, but for
help.

\begin{table}
\begin{center}
  \begin{tabular}{llclcc}  \hline
  Process &  & Number of diagrams & Comment & Check & Reference \\
  \hline\hline
  {\bf SUSY} & {\bf QED} & & & & \\ \hline
$e^- e^- \rightarrow$ & $\selecR^- \selecR^-$ & 2 & Majorana-flip & OK & \\
                    & $\selecL^- \selecL^-$ & 2 & in internal lines & OK & \\
                    & $\selecR^- \selecL^-$ & 2 & & OK & \\  \hline
$e^- e^+ \rightarrow$ & $\selecR^- \selecR^+$ & 2 & Including pair & OK &
		    \cite{mj}\\
        & $\selecL^- \selecL^+$ & 2 & annihilation & OK & \cite{mj} \\ \hline
$e^- e^+ \rightarrow$ & $\selecR^- \selecL^+$ & 1 & Values are & OK &
		    \cite{mj} \\
        & $\selecR^+ \selecL^-$ & 1 & equal & OK & \cite{mj} \\  \hline
$e^- e^+ \rightarrow$ & $\photino \photino$ & 4 & F-B symmetric & OK &
	\\  \hline
$e^- e^+ \rightarrow$ & $\photino \photino \gamma~$ & 12 & Final 3-body
   & OK & \cite{tk} \\ \hline\hline
  {\bf MSSM} & & & & \\ \hline
$e^- e^- \rightarrow$ & $\selecL^- \selecL^-$ & 8 & 4 Majorana fermions & OK &
   \\ \hline
$e^- e^+ \rightarrow$ & $\charglm \charglp$ & 3 & & OK &\\  \hline
  \end{tabular}
\end{center}
  \hspace*{4cm} Table~I. The list of the tested processes.
\end{table}%

\quad\\
\noindent{\bf Summary}

    We introduce a new method to treat Majorana fermions on the GRACE system
for the automatic computation of the matrix elements for the processes of the
SUSY models.  In the first instance, we have constructed the system for the
processes of the SUSY QED because we should test our algorithm with the
simplest case.  The numerical results convince us that our algorithm is
correct.

    It is remarkable that our system is also applicable to another model
including Majorana fermions ({\it e.g.} MSSM) once the definition of the model
file is given.  We should computate the single-photon event from $e^- e^+
\rightarrow \lsp \lsp \gamma$~\cite{tk}, and the resultant single-positron
(electron) event from the single-selectron production $e^- e^+ \rightarrow
\selecmp \lsp e^\pm$~\cite{spu} as soon as possible.  It should be emphasized
that the GRACE system including SUSY particles is the powerful tool for the
purpose.

\quad\\
\noindent{\bf Acknowledgements}

  This work was supported in part by the Ministry of Education,
  Science and Culture, Japan under Grant-in-Aid for International
  Scientific Research Program No.04044158.  One of the author (M.J.) and Dr.
  H. Tanaka have been also indebted to the above-mentioned Ministry under
  Grant-in-Aid No.06640411.

\end{document}